\documentclass[aps,amsfonts,nofootinbib,twocolumn]{revtex4}
\usepackage{graphicx}
\usepackage{amsmath}
\def\be{\begin{equation}}
\def\ee{\end{equation}}
\def\bea{\begin{eqnarray}}
\def\eea{\end{eqnarray}}

\begin{document}

\title{Structure formation in the DGP cosmological model}

\author{Kazuya Koyama, Roy Maartens}

\affiliation{\vspace*{0.2cm} Institute of Cosmology \&
Gravitation, University of Portsmouth, Portsmouth~PO1~2EG, UK
\vspace*{0.2cm}}

\date{\today}

\begin{abstract}

The DGP brane-world model provides an alternative to the standard
LCDM cosmology, in which the late universe accelerates due to a
modification of gravity rather than vacuum energy. The
cosmological constant $\Lambda$ in LCDM is replaced by a single
parameter, the crossover scale $r_c$, in DGP. The Supernova
redshift observations can be fitted by both models, with
$\Lambda\sim H_0^2$ and $r_c \sim H_0^{-1}$. This degeneracy is
broken by structure formation, which is suppressed in different
ways in the two models. There is some confusion in the literature
about how the standard linear growth factor is modified in DGP.
While the luminosity distance can be computed purely from the
modified 4-dimensional Friedman equation, the evolution of density
perturbations requires an analysis of the 5-dimensional
gravitational field. We show that if the 5-dimensional effects are
inappropriately neglected, then the 4-dimensional Bianchi identities 
are violated and the computed growth factor is incorrect. By using the
5-dimensional equations, we derive the correct growth factor.

\end{abstract}

\maketitle

\section{Introduction}

The acceleration of the late-time universe, as implied by
observations of Supernovae redshifts, cosmic microwave background
anisotropies and the large-scale structure, poses one of the
deepest theoretical problems facing cosmology. Within the
framework of general relativity, the acceleration must originate
from a dark energy field with effectively negative pressure, such
as vacuum energy or a slow-rolling scalar field (``quintessence").
So far, none of the available models has a natural explanation.
For example, in the simplest option of vacuum energy, leading to
the ``standard" LCDM model, the incredibly small,
\begin{equation}
\rho_{\Lambda,\mbox{obs}}={\Lambda \over 8\pi G}\sim H_0^2M_P^2
\ll \rho_{\Lambda,\mbox{theory}}\,,
\end{equation}
and incredibly fine-tuned,
\begin{equation}
\Omega_{\Lambda}\sim \Omega_m\,,
\end{equation}
value of the cosmological constant cannot be explained by current
particle physics.

An alternative to dark energy plus general relativity is provided
by models where the acceleration is due to modifications of
gravity on very large scales, $r\gtrsim H_0^{-1}$. One of the
simplest covariant models is based on the Dvali-Gabadadze-Porrati
(DGP) brane-world model \cite{Dvali:2000rv}, in which gravity leaks off the
4-dimensional Minkowski brane into the 5-dimensional ``bulk"
Minkowski spacetime at large scales. At small scales, gravity is
effectively bound to the brane and 4-dimensional Newtonian 
dynamics is recovered to a good approximation. The transition
from 4- to 5-dimensional behaviour is governed by a crossover
scale $r_c$; the weak-field gravitational potential behaves as
\begin{equation}
\Psi \sim \left\{ \begin{array}{lll} r^{-1} & \mbox{for} & r< r_c
\\ r^{-2} & \mbox{for} & r> r_c \end{array}\right.
\end{equation}
The DGP model was generalized by Deffayet to a
Friedman-Robertson-Walker brane in a Minkowski bulk \cite{Deffayet:2000uy}; 
the gravity leakage at late times initiates acceleration -- not due to any
negative pressure field, but due to the weakening of gravity on
the brane. The energy conservation equation remains the same as in
general relativity, but the Friedman equation is modified:
\begin{eqnarray}
&& \dot\rho+3H(\rho+p)=0\,,\label{ec} \\ && H^2-{H \over r_c}=
{8\pi G \over 3}\rho\,. \label{f}
\label{Fried}
\end{eqnarray}
It is important to stress that the modification to the Friedman
equation is derived from a covariant 5-dimensional action and
junction conditions across the brane \cite{Deffayet:2000uy}.

The modified Friedman equation~(\ref{f}) shows that at late times
in a CDM universe, with $\rho\propto a^{-3}\to0$, we have
\begin{equation}
H\to H_\infty= {1\over r_c}\,.
\end{equation}
Since $H_0>H_\infty$, in order to achieve acceleration at late
times, we require $r_c\gtrsim H_0^{-1}$, and this is confirmed by
fitting SN observations \cite{Deffayet:2002sp}.

Like the LCDM model, the DGP model is simple, with a single
parameter $r_c$ to control the late-time acceleration. The DGP
model does not provide a natural solution to the late-acceleration
problem; similarly to the LCDM model, where $\Lambda$ must be
fine-tuned, the DGP parameter $r_c$ must be fine-tuned to match
observation. 
Furthermore, it has been recognized that the DGP model 
suffers from potentially serious theoretical problems, such as the 
existence of a ghost in de Sitter solutions of Eq.~(\ref{Fried}) 
and the strong coupling problem \cite{problems}. 
Nevertheless, like LCDM, the DGP is a simple covariant model that
makes testable predictions, in particular the prediction of
late-time acceleration. (It is worth noting that the original DGP
model with a Minkowski brane was not introduced to explain
acceleration -- the generalization to a Friedman brane was
subsequently found to be self-accelerating.)

LCDM and DGP can both account for the SN observations, with
$\Lambda\sim H_0^2$ and $r_c\sim H_0^{-1}$ respectively. This
degeneracy may be broken by observations based on structure
formation, since the two models predict different suppression of
density perturbations. For LCDM, the analysis of density
perturbations is well understood. For DGP it is much more subtle
and complicated; although matter is confined to the 4-dimensional
brane, gravity is fundamentally 5-dimensional, and the bulk
gravitational field responds to and backreacts on density
perturbations. The background cosmological dynamics, and hence the
predicted SN redshifts, can be treated purely via the modified
4-dimensional Friedman equation. By contrast, the evolution of
density perturbations requires an analysis based on the
5-dimensional nature of gravity.

\section{Perturbation equations on the brane}

Effective covariant equations on the brane, for an arbitrary brane
metric and matter distribution, can be derived by projecting the
5-dimensional Einstein equations ($G^{(5)}_{ab}=0$) and applying
the Israel junction conditions with reflection symmetry at the
brane \cite{Shiromizu:1999wj, Maeda:2003ar}
\begin{equation}
G_{\mu \nu} =(16\pi G r_c)^2 \Pi_{\mu
\nu}-{ E}_{\mu \nu}\,, \label{1-1-1}
\end{equation}
where
\begin{eqnarray}
\tilde{T}_{\mu \nu} &=& T_{\mu \nu} - (8\pi G)^{-1} G_{\mu \nu}\,,
\\ \Pi_{\mu \nu}&=& -\frac{1}{4} \tilde{T}_{\mu \alpha}
\tilde{T}_{\nu}^{\alpha} +\frac{1}{12} \tilde{T}^{\alpha}_{\alpha}
\tilde{T}_{\mu \nu}\nonumber\\&&~{}+ \frac{1}{24}\left[3
\tilde{T}_{\alpha \beta} \tilde{T}^{\alpha \beta}-
(\tilde{T}^{\alpha}_{\alpha})^2\right] g_{\mu \nu}\,,
\end{eqnarray}
and ${ E}_{\mu \nu}$ is the trace-free projection of the 5D Weyl
tensor. The energy-momentum tensor satisfies
\begin{equation}
\nabla^{\mu} T_{\mu \nu}=0,
\end{equation}
as in general relativity.
The 4D Bianchi identity
imposes the constraint
\begin{equation}\label{bi}
\nabla^{\mu} E_{\mu \nu} = (16\pi G r_c)^2 \nabla^{\mu} \Pi_{\mu
\nu}\,.
\end{equation}

In the background spacetime, $E_{\mu \nu}=0$ and Eq.~(\ref{1-1-1})
implies the modified Friedman equation~(\ref{f}). The perturbed
FRW brane has a nonzero $E_{\mu \nu}$, which encodes the effects
on the brane of the bulk gravitational field. The perturbed 5D
field equations are needed to determine the evolution of $\delta
E_{\mu \nu}$. We can parametrize the (gauge-invariant) scalar
perturbations of $E_{\mu \nu}$ as an effective fluid, with density
perturbation $\delta\rho_E$, isotropic pressure perturbation
${1\over3}\delta \rho_{{ E}}$, anisotropic stress perturbation
$\delta\pi_E$ and energy flux perturbation $\delta q_E$ (see the
appendix). Scalar metric perturbations are given in longitudinal
gauge by
\begin{equation}\label{smp}
ds^2= -(1+2 \Psi)dt^2 + a^2 (1+2 \Phi) d\vec{x}^2\,,
\end{equation}
and the perturbed energy-momentum tensor for matter is given by 
\begin{equation}
\delta T^{\mu}_{\nu} = 
\left(
\begin{array}{cc}
 -\delta \rho & a \delta q_{,i} \\
 -a^{-1} \delta q^{,i}  & \delta p \: 
 \delta^{i}_{\:\: j} \\
\end{array}
\right).
\end{equation}

The Bianchi identity~(\ref{bi}) gives evolution equations~(\ref{bianchit}) and 
(\ref{bianchii}) for $\delta \rho_E$ and $\delta q_E$. The perturbed 4D field equations (\ref{1-1-1})
imply a modified Poisson equation (\ref{einstein-tf}) and constraint equation
(\ref{einstein-ijt}). The 4D perturbation equations are not closed, due to the presence
of Weyl anisotropic stress $\delta\pi_E$. The 5D perturbation
equations are needed to determine this stress. This is an
extremely complicated problem in full generality, and especially
on large scales where the 5D effects become strong. For the
structure formation problem, we focus on subhorizon scales, with
$k/a\ll H^{-1}$ and $k/a\ll r_c$, where $k$ is the comoving wavenumber. 
In addition, we use the quasi-static
approximation which is relevant for structure formation and neglect
time-derivatives terms relative to gradient terms. 
Then Eq.~(\ref{bianchit}) implies
\begin{equation}
\delta q_E =0.
\end{equation}

With $\delta q_E= 0$, the key perturbation equations reduce to
\begin{eqnarray}
&&\frac{k^2}{a^2} \Phi = 4\pi G \left( \frac{2 H r_c}{ 2 Hr_c -1}
\right) \left (\rho \bigtriangleup - \frac{\delta \rho_{ E}}{2
Hr_c} \right), \label{einstein-t}\\
&&\Phi+\Psi = 8\pi G\, \frac{H}{r_c(\dot{H} + 2 H^2)-H}\, a^2 \delta
\pi_{ E}\,, \label{einstein-ijt}\\
&& k^2 \delta \pi_{ E} - \frac{1}{2} \delta \rho_{ E} = 
r_c \frac{\dot{H}}{H} F,\label{bianchi}\\
&& \ddot{\bigtriangleup}+2 H \bigtriangleup = - {k^2\over a^2}
\Psi\,, \label{matter}
\end{eqnarray}
where
\begin{equation}
F \equiv \frac{\rho \bigtriangleup}{1-2Hr_c} +
\frac{H}{r_c(2 H^2+\dot{H}) -H} k^2 \delta \pi_{ E} + \frac{\delta
\rho_{ E}}{2 Hr_c -1} \!,
\end{equation}
and the (gauge-invariant) comoving density perturbation is
\begin{equation}\label{cdp}
\rho\Delta=\delta\rho-3H a \delta q\,.
\end{equation}

The equations~(\ref{einstein-t})--(\ref{matter}) are not closed
since $ \delta E_{\mu \nu}$ is not determined by purely 4D
equations. The ad hoc assumption $ \delta E_{\mu \nu}=0$, i.e.,
\begin{equation}
\delta \rho_{{ E}} = \delta \pi_{ E} =0\,,
\end{equation}
has been effectively adopted in some papers \cite{wrong}. However, it is
manifest that this assumption is {\em not} consistent with the
Bianchi identity Eq.~(\ref{bianchi}). Thus the analysis based on
this assumption must be revisited.

Within our approximations, the Bianchi identity suggests that
there exists a family of solutions characterized by
\begin{equation}\label{c}
\delta \rho_{{ E}} = C(Hr_c) k^2 \delta \pi_{ E}\,,
\end{equation}
where $C(Hr_c)$ is some function of $Hr_c$. The correct $C(Hr_c)$
can be obtained only when we properly consider the 5D equations of
motion.

\section{The 5-dimensional equations}

The 5D background Minkowski metric is given by
\begin{equation}
ds_{(5)}^2=  - n(y,t)^2 dt^2 +b(y,t)^2 d\vec{x}^2+dy^2\,,
\end{equation}
where the brane is at $y=0$ and
\begin{equation}
b=a(1+H y)\,,~~ n= 1 + \left(\frac{\dot{H}}{H} + H \right) y\,.
\end{equation}
Because the background bulk is just a Minkowski spacetime, we can
easily find a Mukohyama master variable~ $\Omega(t,\vec x,y)$ 
\cite{Mukohyama:2000ui},
from which all the solutions for the metric perturbations can be
constructed (see the appendix). In the quasi-static approximation
relevant for structure formation, we can neglect $\dot\Omega$
relative to $\Omega'$, and the 5D wave equation for $\Omega$
becomes
\begin{equation}\label{mv}
\Omega'' + \left(- 3 \frac{b'}{b} + \frac{n'}{n}\right) \Omega' -
\frac{k^2}{b^2} \Omega =0\,,
\end{equation}
where a dot denotes a derivative with respect to time $t$ and a prime
a derivative with respect to $y$.
In the quasi-static approximation, the solution is
\begin{equation}\label{mvs}
\Omega =c_1 \left[ (1+H y)^{-k/aH} + c_2 (1+H y)^{k/aH} \right].
\end{equation}
Regularity of the bulk perturbations as $y\to\infty$ requires that
$c_2=0$.

Perturbations of the Weyl fluid $\delta E_{\mu \nu}$ can be
expressed in terms of $\Omega$,
\begin{eqnarray}
8\pi G \delta \rho_{{E}} &=& \frac{k^4 }{3
a^5}\Omega\big|_{y=0}\,, \label{rhoe}\\
8\pi G \delta \pi_{{E}} &=& \frac{1}{6 a^3} \left(\frac{k^2}{a^2}
\Omega - \frac{3 \dot{H}}{H} \Omega' \right)_{y=0}. \label{pie}
\end{eqnarray}
By Eq.~(\ref{bound}) with $c_2=0$, we can neglect the last term on
the right of Eq.~(\ref{pie}). It follows that
\begin{equation}
\label{eom}
\delta \rho_{{E}} =2 k^2 \delta \pi_{ E}\,,
\end{equation}
and we have determined the function $C$ in Eq.~(\ref{c}) as $C=2$.

\section{The growth factor}
In the appendix, we find the solution for $\Omega$ 
(\ref{solom}) by imposing the perturbed junction condition (\ref{junction}).  
Then using Eqs.~(\ref{A}), (\ref{R}), (\ref{metricphi}), (\ref{metricpsi})
and (\ref{bound}),  
the solutions for the brane metric perturbations
are
\begin{eqnarray}
\frac{k^2}{a^2} \Phi &=& 4\pi G \left(1- \frac{1}{3 \beta} \right)
\rho \triangle \label{solphi}\\
\frac{k^2}{a^2} \Psi &=& -4\pi G \left(1 + \frac{1}{3 \beta}
\right) \rho \triangle \label{solpsi}
\end{eqnarray}
where
\begin{equation}
\beta =1 -2 r_c H \left(1+ \frac{\dot{H}}{3 H^2} \right).
\end{equation}
This agrees with the results obtained by Lue, Scoccimarro and Starkman
\cite{Lue:2004rj}. They find spherical symmetric solutions 
by closing the 4D equations using an anzatz for the metric and checked
in retrospect that the obtained solutions satisfy the regularity in the 
bulk. Here we have shown that the solutions (\ref{solphi}) and (\ref{solpsi}) 
are uniquely determined by the regularity condition 
in the bulk within our approximations. 

We can check that our solutions are consistent with the Bianchi identity. 
The Bianchi identity Eq.~(\ref{bianchi}) gives non-trivial equations 
for $\delta \rho_{E}$ and $\delta \pi_{ E}$ with the condition (\ref{eom}) and 
they can be solved 
in terms of $\rho \bigtriangleup$. Then substituting the solutions into
Eqs.~(\ref{einstein-t}) and (\ref{einstein-ijt}), we arrive at the
solutions for $\Phi$ and $\Psi$ as in Eqs.~(\ref{solphi}) and
(\ref{solpsi}). 


The modified Poisson equation~(\ref{solphi}) shows the suppression
of growth due to gravity leakage. The rate of growth is determined
by $\Delta$, and for CDM,
\begin{equation}
\ddot{\bigtriangleup} + 2 H \dot{\bigtriangleup}=-\frac{k^2}{a^2}
\Psi\,.
\end{equation}
which leads to
\begin{equation}\label{dpe}
\ddot{\bigtriangleup} + 2 H \dot{\bigtriangleup}=4\pi G \left(1 +
\frac{1}{3 \beta} \right) \rho \bigtriangleup\,.
\end{equation}
Thus the growth rate receives an additional modification from the
time variation of Newton's constant through $\beta$. 
This effect can be described by a linearized scalar-tensor gravity with 
Brans-Dicke parameter \cite{Lue:2004rj}
\be
\omega=\frac{3}{2}(\beta-1),
\ee
where the gravitational scalar corresponds to the bending of the brane
Eq.~(\ref{bending}).

In Fig.~1, we show the linear growth factor $\Delta/a$ for the DGP
model, and compare it with LCDM, with the incorrect DGP result (in
which the inconsistent assumption $\delta E_{\mu\nu}=0$ is
effectively adopted), and with the general relativity dark energy
model whose background evolution matches that of the DGP model 
(see Fig.~2).

\begin{figure}[t]
\centerline{
\includegraphics[width=8cm]{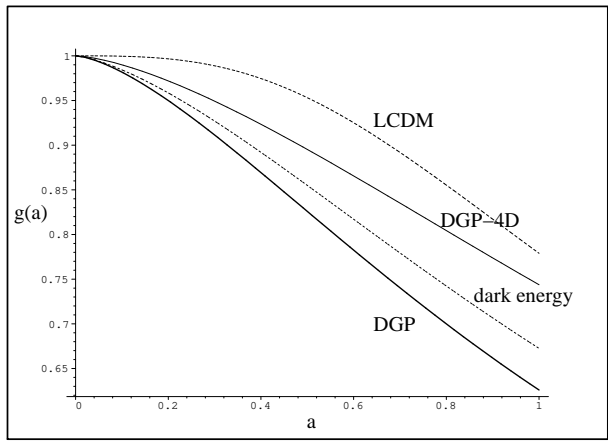}}
\caption{The growth history $g(a)=\bigtriangleup(a)/a$ is shown for 
LCDM (long dashed) and DGP (solid, thick). The growth history for a 
dark energy model (short dashed) with the identical expansion histories 
with DGP is also shown (see Fig.~2) \cite{Linder}. Due to the time variation of Newton's 
constant through $\beta$ in Eq.~(\ref{dpe}), the growth factor $g(a)$ receives 
an additional suppression compared with the dark energy model. 
DGP-4D (solid, thin) shows the incorrect 
result in which the inconsistent assumption $\delta E_{\mu \nu}=0$ 
is adopted. We set the desity parameter for matter today as 
$\Omega_{m0}=0.3$.}
\label{fig:fig1}
\end{figure}

\section{conclusion}
We have shown by a careful analysis of the 5D perturbation
equations that subhorizon density perturbations in the DGP
cosmological models evolve according to Eq.~(\ref{dpe}), that the
modified Poisson equation is given by Eq.~(\ref{solphi}), and that
the gravitational potential is given by Eq.~(\ref{solpsi}). These
expressions obey the constraint imposed by the 4D Bianchi
identity, and they correct those expressions in the literature \cite{wrong}
which violate the Bianchi identity. Our results confirm the
results derived via another approach by Lue, Scoccimarro and Starkman
\cite{Lue:2004rj}.
Although they also used the 5D equations to check consistency of
their ansatz, they were not able to show, as we have done here,
that their results follow uniquely from the 5D equations.

The correct equations for subhorizon density perturbations are
crucial for meaningful tests of DGP predictions against structure
formation observations. And such tests are essential for breaking
the degeneracy with LCDM that arises with SN redshift
observations \cite{wrong, Linder}. 
The distance-based SN observations draw only upon
the background 4D Friedman equation~(\ref{f}) in DGP models, and
therefore there are quintessence models in general relativity that
can produce precisely the same SN redshifts as DGP (see Fig.~2). By
contrast, structure formation observations require the 5D
perturbations in DGP, and one cannot find equivalent general
relativity models. Previous analyses of structure formation that
are based on the incorrect equations for density perturbations \cite{wrong}
will need to be re-visited.

\begin{figure}[t]
\centerline{
\includegraphics[width=8cm]{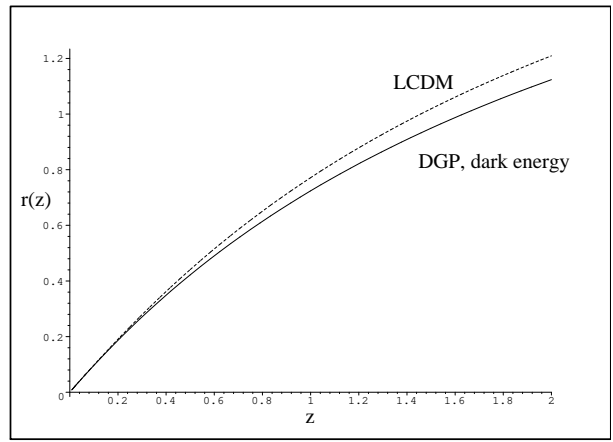}}
\caption{The expansion history $r(z)=\int^z_0 dz H(z)^{-1}$ is shown for LCDM (dashed)
and DGP (solid) where $1+z=a^{-1}$. Expansion history for DGP can be mimicked by a dark energy 
model with the equation of state $w(a)=w_0 +w_a (1-a)$ with $w_0=-0.78$ and $w_a=0.32$
\cite{Linder}.}
\label{fig:fig2}
\end{figure}

The linear growth factor, as shown in Fig.1, is the basis for
tests of DGP against structure formation observations. But one
also needs the metric perturbations. It follows from
Eqs.~(\ref{solphi}) and (\ref{solpsi}) that
\begin{eqnarray}
\frac{k^2}{a^2} (\Phi - \Psi) &=& 8\pi G \rho \bigtriangleup\,,
\label{pmp}\\ \frac{k^2}{a^2} (\Phi + \Psi) &=& -{8\pi G\over
3\beta}\, \rho \bigtriangleup \propto \delta\pi_E\,,\label{ppp}
\end{eqnarray}
where the proportionality follows from Eq.~(\ref{einstein-ijt}).
Equation~(\ref{ppp}) differs from general relativity, where the
absence of matter anisotropic stress implies $\Psi=-\Phi$. In the
DGP model, the 5D gravitational field produces an anisotropic
stress on the 4D universe. Equation~(\ref{pmp}) is the same as in
general relativity -- and this equation basically determines the
integrated Sachs-Wolfe and weak lensing effects. 
This means that
the formulas will take the same form as in general relativity, but
the outcome will be different, because $\Delta$ will differ from
the general relativity case.

We have also emphasized that our results 
(and those of Ref.~\cite{Lue:2004rj})
apply on subhorizon scales. On superhorizon scales, where the 5D
effects are strongest, the problem has yet to be solved. Thus a
computation of CMB anisotropies will not be reliable on the
largest scales.

Furthermore, we have restricted our discussion to the linear
regime. In the nonlinear regime, the DGP theory approaches general
relativity \cite{nonlinear}. A full computation of weak lensing will require a
careful matching of the linear DGP to the nonlinear general
relativity regimes.

\[ \]
{\bf Acknowledgements:} KK and RM are supported by PPARC. We thank
Arthur Lue for useful discussions.

\appendix

\section{Perturbation equations}

We can parametrize the scalar perturbations of $E_{\mu \nu}$ as an
effective fluid,
\begin{equation}
\delta { E}^{\mu}_{\,\,\nu} = -8\pi G \left(
\begin{array}{ccc}
-\delta \rho_{{ E}} & & a \delta q_{{ E},i} \\
a^{-1} \delta q_E^{\,\,\,,i} &  & {1\over3}\delta \rho_{{ E}} \:
\delta^{i}_{\,\,j}
+ \delta \pi^{i}_{E \,j}\\
\end{array}
\right),
\end{equation}
where $\delta\pi^E_{ij}=\delta\pi^E_{,ij}-
{1\over3}\delta\pi_{E\,,k}^{,k}\delta_{ij}$. The 4D Bianchi
identity Eq.~(\ref{bi}) imposes
\begin{eqnarray} \label{bianchit}
&& \dot{\delta \rho}_E + 4 H \delta \rho_E - a^{-1} k^2 \delta q_E
=0\,,\\ \label{bianchii}&& \dot{\delta q}_{E} + 4 H \delta q_E + a^{-1}
\left(\frac{1}{3} \delta \rho_E - \frac{2}{3} k^2 \delta \pi_E
\right) \nonumber\\&&{} =-a^{-1} \frac{2}{3} r_c \frac{\dot{H}}{H}
\biggr\{ -\frac{\rho \bigtriangleup}{2 Hr_c -1} + \frac{\delta
\rho_E}{2 Hr_c-1}  \nonumber\\&&{} + \frac{1}{r_c (2H +
{\dot{H}}/{H})-1} k^2 \delta \pi_E \biggl\}. \label{EulerE}
\end{eqnarray}
Combining the $(t,t)$ component and the $(0,i)$ component of the perturbed 4D field 
equations Eq.~(\ref{1-1-1}), we get the modified Poisson equation
\begin{equation}
\frac{k^2}{a^2} \Phi = 4\pi G \left( \frac{2 H r_c}{ 2 Hr_c -1}
\right) \left (\rho \bigtriangleup - \frac{\delta \rho_{ E} -3Ha \delta q_E }{2
Hr_c} \right),
\label{einstein-tf}
\end{equation}
where $\rho \bigtriangleup = \delta \rho - 3 Ha \delta q$. 
The traceless part of $(i,j)$ component gives Eq.~(\ref{einstein-ijt}).
(We follow the standard assumption that the CDM anisotropic stress
vanishes $\delta\pi=0$.)

In order to solve the perturbations in this background, it is
convenient to use 5D Longitudinal gauge, given by
\begin{eqnarray}
ds^2 &=& (1+2 A_{yy})dy^2+nA_{y}dydt -n^2(1+2 A^2) dt^2 \nonumber\\
&& +b^2(1+2 {\cal R})d\vec{x}^2\,.
\end{eqnarray}
In Minkowski spacetime, if the metric perturbations are
derived from a ``master variable'', $\Omega$, 
the perturbed 5D Einstein equations yield a single
wave equation governing the evolution of the master variable
$\Omega$ in the bulk:
\begin{equation}
\label{scalarmastereom}
 - \left( {1\over nb^3} \dot\Omega \right)^{\displaystyle\cdot}
 + \left( {n\over  b^3} \Omega' \right)^\prime
 - {k^2 n \over b^5} \Omega
 = 0
 \,.
\end{equation}
Within our approximation, we can neglect time-derivative terms 
and the metric perturbations are written in terms of $\Omega$ as
\begin{eqnarray}
A &=& -\frac{1}{6b} \left\{ -3\left(\frac{n'}{n}-2 \frac{b'}{b}
\right) \Omega'
+ \frac{2 k^2}{b^2} \Omega \right\}, \label{A}\\
R &=& \frac{1}{6b} \left(3 \frac{b'}{b} \Omega' + \frac{k^2}{b^2}
\Omega \right). \label{R}
\end{eqnarray}

An important feature of the DGP model is the brane-bending mode;
in 5D longitudinal gauge, the location of the brane is perturbed
and given by \cite{Deffayet1}
\begin{equation}
y=\xi=-r_c(\Phi+\Psi).
\label{bending}
\end{equation}
Then the induced metric perturbations on the brane are given by
\begin{equation}
\Psi =A -\left(\frac{\dot{H}}{H}+ H \right) \xi, \quad \Phi=R -H
\xi,
\end{equation}
from a gauge transformation. We can express $\Psi$ and $\Phi$ in
terms of $A$ and $R$ as
\begin{eqnarray}
\Phi &=& \frac{1}{1-r_c \left(\frac{\dot{H}}{H}+2 H \right)} \left
\{ \left(1 - r_c \frac{\dot{H} +H^2}{H} \right) R + Hr_c A
\right\}, \label{metricphi}
\nonumber\\ \\
\Psi &=& \frac{1}{1-r_c \left(\frac{\dot{H}}{H}+2 H \right)} \left
\{ (1-Hr_c) A + r_c \frac{\dot{H} +H^2}{H}  R \right\}.
\nonumber\\
\label{metricpsi}
\end{eqnarray}
The perturbed junction condition is
\begin{equation}
4\pi G\rho \bigtriangleup = \frac{k^2}{a^2} \Phi + \frac{k^2}{2
a^2} \frac{\xi}{r_c} - \frac{1}{4 r_c} \frac{k^2}{a^3}
\Omega'\big|_{y=0}\,. \label{junction}
\end{equation}
This junction condition gives the modified Poisson equation where
the last two terms describe corrections from the 5D gravity. 

Note that $\Omega'$ appears in the expressions for $\delta \rho_{
E}$ and $\delta \pi_{ E}$. Unless we solve the 5D equations and
impose proper boundary conditions, $\Omega$ and $\Omega'$ on the
brane are independent. This means that we can impose arbitrary
conditions for the relation between $\delta \rho_{ E}$ and $\delta
\pi_{ E}$. Indeed by choosing appropriate $c_2$ in
Eq.~(\ref{mvs}), we can have any $C(Hr_c)$ in Eq.~(\ref{c}).
However, as described in the main text, the regularity condition
impose $c_2 =0$. Then we get
\begin{equation}
H \Omega' \ll \frac{k^2}{a^2} \Omega,
\label{bound}
\end{equation}
for $k/aH \gg 1$ and we can neglect the terms proportional to
$\Omega'$. 
However, the Poisson equation Eq.~(\ref{junction}) still receives a correction 
from the brane bending mode $\xi$ and we get a 
linearized scalar-tensor gravity \cite{Deffayet2}. 
The junction condition Eq.(\ref{junction}) gives
\begin{equation}
\frac{k^4 \Omega}{2 a^5} =\frac{1- 2 H r_c (1 + {\dot{H}}/{2H^2}
)} {1- 2 H r_c  (1 + {\dot{H}}/{3H^2} )} \kappa_4^2 \rho
\bigtriangleup.
\label{solom}
\end{equation}
Then from Eqs.~(\ref{A}), (\ref{R}), (\ref{metricphi}) and
(\ref{metricpsi}), the solutions for metric perturbations can be
obtained as Eqs.~(\ref{solphi}) and (\ref{solpsi}). We can also
derive the solutions for $\delta \rho_E$ and $\delta \pi_E$ from
Eqs.~(\ref{rhoe}) and (\ref{pie}) and show that they are
consistent with the 4D Bianchi identity.

\end{document}